\documentclass[preprint,12pt]{elsarticle}  

\usepackage{graphicx}
\graphicspath{{./figures/}{./inkscape/}{../figures/}}
\journal{}
\usepackage{amsmath,amsfonts,amsthm,bm,mathtools} 
\usepackage{amssymb} 
\usepackage{lipsum}
\usepackage{float}
\usepackage{pgfplots}
\usepackage{graphicx,amsmath,upgreek,bm}
\usepackage{subcaption}
\usepackage{siunitx}
\usepackage{makecell}
\usepackage{booktabs}
\pgfplotsset{compat=newest}
\usepackage{multicol}
\usepackage{xcolor,colortbl}
\pgfplotsset{plot coordinates/math parser=false}
\usepackage{multicol} 
\usepackage{amsthm}

\usepackage{geometry}
\usepackage{afterpage}
\usepackage{longtable}
\usepackage[tableposition=top]{caption}
\usepackage{multirow}
\usepackage[bb=boondox]{mathalpha}
\definecolor{label}{rgb}{0.1,0.25,.65}
\definecolor{title}{rgb}{0.1,0.25,.65}
\definecolor{label}{rgb}{0,0,0}
\definecolor{title}{rgb}{0,0,0}
\definecolor{cite}{rgb}{0.6,0.1,.2}
\usepackage[
font=normalsize,
labelfont={bf,color=label}, 
justification=justified,
format=plain]{caption} 
\usepackage{subcaption}
\newlength\fwidth
\usepackage{natbib}
\usepackage{bibentry}
\definecolor{best_acc}{rgb}{0,0.5,0}
\definecolor{refcol}{rgb}{.25,0,1}
\definecolor{mygray}{gray}{0.85}
\definecolor{mycite}{gray}{0.0}
\let\oldbibliography\thebibliography
\renewcommand{\thebibliography}[1]{%
	\oldbibliography{#1}%
	\setlength{\itemsep}{0pt}%
}
\definecolor{mygray}{gray}{0.85}
\definecolor{edit1}{rgb}{0,0,0}
\definecolor{edit2}{rgb}{1,0,0}

\usepackage[misc]{ifsym}
\usepackage{fontawesome}
\usepackage{ragged2e}
\title{\Large{\textbf{Spectral Characterization of Wave Scattering at a Granular–Elastic Solid Interface: From Hyperbolic Wave Propagation to Near-Parabolic Diffusion}}}

\author[inst1]{\textcolor{title}{Joshua R.~Tempelman}}
\author[inst2]{\textcolor{title}{Chongan Wang}}
\author[inst3]{\textcolor{title}{Alexander F.~Vakakis}}

\affiliation[inst1]{organization={\textcolor{title}{Space Remote Sensing and Data Science Group,  Los Alamos National Laboratory, Los Alamos NM, USA}
}}

\affiliation[inst2]{organization={\textcolor{title}{G.W. Woodruff School of Mechanical Engineering, Georgia Institute of Technology}
}}
\affiliation[inst3]{organization={\textcolor{title}{Department of Mechanical Science and Engineering, University of Illinois Urbana-Champaign}
}}

\usepackage{titlesec}
\titleformat{\section}
{
	\bfseries
	\large
	\color{label}
}  
{\thesection}  
{1em}  
{}

\titleformat{\subsection}
{
	\bfseries
	\normalsize
	\color{label}
}  
{\thesubsection}
{1em}
{}
\usepackage{todonotes}
\usepackage{lineno}

\usepackage{setspace}
\newcommand{\rev}[1]{{\textcolor{edit1}{#1}}}

\usepackage{mathrsfs}

\usepackage[colorlinks=true,
linkcolor=cite,
citecolor=cite,
urlcolor=cite]{hyperref} 

\begin{document}

	
	\begin{abstract}
		We present a method based on acoustic wavenumber imaging algorithms to quantify the spectral content of strongly nonlinear energy scattering of a propagating wavefront across the discrete-continuum interface of a 2D hybrid system composed of an ordered granular layer in contact with a thin elastic plate. We consider snapshots of the transmitted wavefront at given time instants, which are filtered across the wavenumber domain by applying the spatial Fourier Transform (FT), and then the filtered wavefields are transformed back to the spatial domain by inverse spatial FT. This yields a spectral decomposition of the given snapshots at varying center wavenumbers.	Based on this postprocessing method, the scattering of the kinetic energy in the receiving medium (plate) can be studied in the wavenumber-time domain, proving a quantitative measure of the nonlinear scattering of the transmitted wavefront by the strongly nonlinear 2D granular layer.This postprocessing method enables the detailed quantitative study of the  scattering and spectral energy redistribution of propagating wavepackets in elastic media with embedded linear or nonlinear layers or inclusions. In addition, we show that the spectral evolution of a propagating wavefront past a granular interface exhibits diffusion-like behavior in the wavenumber domain, drawing an analogy between parabolic heat diffusion and classical hyperbolic elastodynamic energy transport.
	\end{abstract}
	
	\maketitle
	
	\section{Introduction}

Ordered granular media are spatially ordered assemblies of interacting viscoelastic granules, 
typically of spherical shape. This class of nonlinear systems has unique features and broad 
applications, ranging from solid state physics and crystal lattices to soil mechanics, 
geophysics, vibrations, and acoustics (e.g., \cite{Nesterenko1994,Nesterenko2001,Daraio2006,Porter2008,Yang2011,Leonard2014,Szelengowicz2013,Lawney2014,Manjunath2014}). 
At the same time, this is a challenging field of study due to the highly discontinuous and 
strongly nonlinear nature of granular interactions, including Hertzian contacts, frictional 
effects, and granular separations and collisions. As an indication of the resulting complexity, 
it is well-known that ordered or disordered granular media under external stimulants may exhibit 
all three states of matter -- solids, liquids, and gases \cite{Andreotti2013}. The challenge then 
is to find ways to incorporate and engineer such highly complex and discontinuous media into 
practical designs of drastically enhanced functionality and performance.

In a previous work \cite{Wang2021a}, wave scattering, i.e., transmission and reflection, 
across the 2D granular--elastic solid interface (also referred to as a hybrid system due to 
its dual discrete and continuum composition) depicted in Fig.~\ref{fig:1} was computationally studied 
by developing a robust iterative--interpolative computational algorithm with adaptive 
(variable) time step to avoid numerical instability; additional works with the same type of 
hybrid systems followed \cite{Wang2021b,Wang2022}. The specific hybrid system studied in 
\cite{Wang2021a} and considered in this work is composed of a 
$0.07\text{m}\times0.14\text{m}\times0.005\text{m}$ rectangular, linearly elastic, thin steel 
plate which is in contact with an initially uncompressed hexagonally ordered granular assembly 
composed of a total of 14 identical, spherical, viscoelastic granules. A uniform shock is applied in the $x$-direction on the 
left (traction-free) boundary of the granular assembly, while the other boundary conditions 
are rigid (clamped) except for the granular--plate interface, and the right boundary of the 
plate which is assumed to be traction-free -- see Fig.~\ref{fig:1}b. The granular medium is modeled by 
discrete elements (DE), and the elastic plate by finite elements (FE) under the assumption of 
plane stress, yielding planar (2D) deformation and velocity fields. The sources of strong and 
non-smooth nonlinearity in this hybrid interface are the granule--granule and granule--plate 
interactions in the form of Hertzian contacts, granular separations and collisions, and 
friction due to granular rotations. Furthermore, the linear steel plate has relatively small 
inherent structural damping, which is neglected altogether, as it contrasts with the highly 
dissipative (and dispersive) nonlinear granular assembly originating from its discrete nature. 
To model friction in the granular assembly, the smooth Coulomb--tanh friction model 
\cite{Pennestri2016} is employed since the discontinuous (classical) Coulomb friction model 
yields numerical instability and prevents convergence of the transient simulations 
\cite{Wang2021a}.

\rev{
Whereas previous research focused largely on the slowing of wave transmission and energy transfer, this work sheds light on the spectral properties of the nonlinear energy scattering at a 2D continuum-discrete material interface.}
Namely, we compare the wavenumber decomposition of wave transmission through a  monolithic 2D interface (e.g., a 2D interface between two linear elastic continua) and a granule-elastic interface (which is commonly referred to as bead-plate interface in this work).
Our goal is to develop a quantitative framework for tracking the dispersion of wave energy in the two systems in the wavenumber domain.
While Ref~\cite{Tempelman2023} has explored the spatial-spectral properties of scattering in 1D phononic media, an extension to 2D wave scattering across granules is still lacking. 
\rev{This work directly addresses this topic by developing a structured methodology for interpreting the spectral composition of energy propagating in elastic continua.}

\rev{We draw inspiration from the domain of guided wave imaging~\cite{Chia2023} which has shown great utility for imaging anomalies and defects in structural elements~\cite{Michaels2009,Rogge2013,Flynn2013,Jeon2017,Jeon2020,Ruzzene2007,Kudela2015}.
However, the utilization of such imaging 
for understanding the spectral make-up of waves scattering across engineered media (e.g., a granule interface) is yet to be explored to the best of the author's knowledge. In this work, we leverage similar spectral characterization schemes  to those used in the NDE community to relate the slowing of energy transfer enabled by a granular medium to the rapid redistribution of spectral energy in the receiving elastic plate.}
\rev{Moreover, by comparing the spectral evolution of the transmitted wavefront past its strong nonlinear dispersion by the granular interface to that of a classical heat diffusion process, we show that the granular medium induces a drastic change in the wave transmission, transforming it from an (typical) hyperbolic type to  =a near-parabolic diffusive response in the receiving elastic continuum.
While it is well established that multiple scattering leads to diffusive transport in the short-wavelength limit~\cite{stam1995multiple,jia2004codalike}, particularly within the framework of radiative transport theory (RTE) applied to Coda waves or porous materials~\cite{mayor2014sensitivity,fuqiang2022progress}, most analogies between RTE and classical wave scattering fundamentally rely on linear analyses of the collective response of a large ensemble of scatterers~\cite{van1999multiple}.
In contrast, our study focuses on a relatively compact region of strongly nonlinear ordered granular scatterers and examines the resulting spectral characteristics of the wave transmission within a fully hyperbolic, continuum-scale receiver (thin elastic plate).
}

After briefly reviewing in Section~\ref{SEC2DSim} the wave scattering across the discrete--continuum 
interface of the hybrid system of Fig.~\ref{fig:1}, we discuss a postprocessing framework for quantifying 
the nonlinear scattering of the primary wave transmitted to the plate following its strong 
dispersion through the granular assembly (Section~\ref{SECProcessing}). This is followed in Section~\ref{SECresults} by the 
application of the postprocessing framework to the quantification of the nonlinear scattering 
of the energy of the transmitted wave at different wavenumber bands, as well as a comparison 
to the case when the hybrid system is replaced by a “monolithic” one, i.e., a system of similar 
geometric dimensions and boundary conditions, but composed of uniform plate material. 
\rev{Section~\ref{SEC:Diff} presents an empirical study comparing the two elastic simulations to a classical classical heat diffusion process.}
Lastly, 
in Section~\ref{SECConclusion} we provide a summary of the main findings of this work and discuss potential 
further work and applications.

	\section{Wave transmission across the 2D granular-elastic solid interface}\label{SEC2DSim}
\begin{figure}[ht]
		\centering
		\includegraphics[width=.75\linewidth]{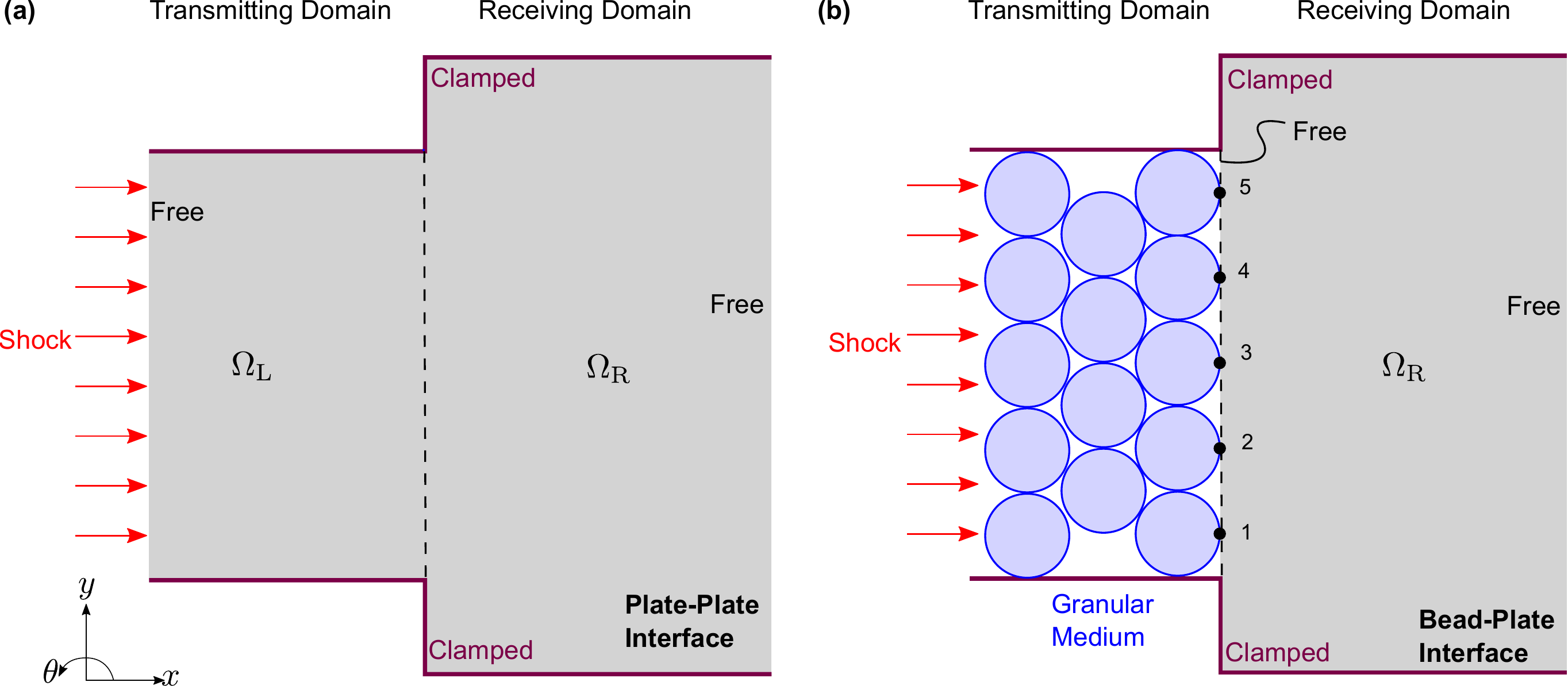}  
		\centering
		\caption{\justifying Schematic of the (a) 2D monolithic system, and (b) 2D hybrid system with boundary conditions.}
		\label{fig:1}
	\end{figure}

	A numerical iterative-interpolative algorithm for computing the shock response of the hybrid system of Fig.~\ref{fig:1}b was developed in~\cite{Wang2021a}, where the challenges associated with ensuring robustness and accuracy of the numerical results were discussed in detail. We only mention at this point that numerical stability issues associated with frictional effects due to granular rotations, granular separations and collisions, and Hertzian contacts had to be carefully addressed by introducing an adaptive time step in the algorithm, which was informed by continuously monitoring the linearized eigenvalues of local 2D maps governing the iterations for the tractions at the granular-elastic solid contact points within each time step. In that way, numerical convergence of the computation at each time step of the simulation was ensured, and the shock response of the 2D hybrid system could be accomplished accurately.

	\begin{figure*}[ht]
		\centering 
		\begin{subfigure}{.5\linewidth}
			\includegraphics[width=\columnwidth]{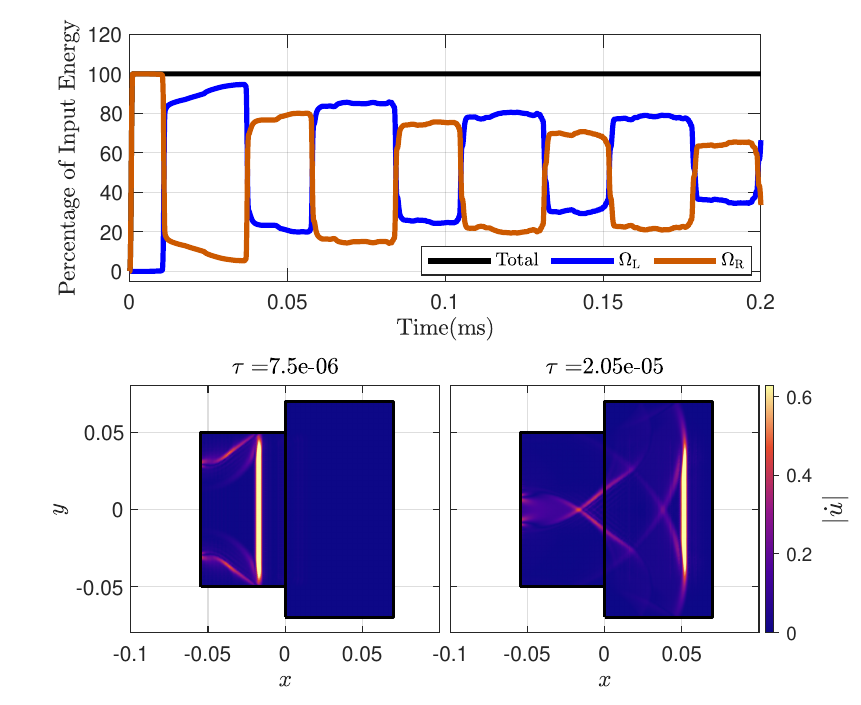}
			\caption{Monolithic-interface system}
		\end{subfigure}%
		\begin{subfigure}{.5\linewidth}
			\includegraphics[width=\columnwidth]{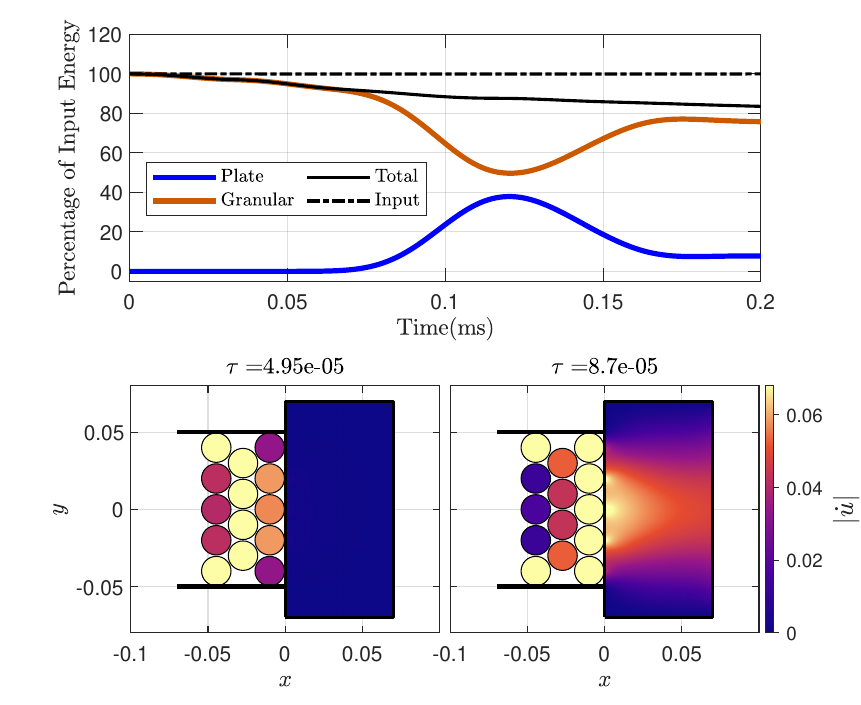}
			\caption{Hybrid granular-interface system}
		\end{subfigure}
		\caption{\justifying Energy measures of the shock responses for the (a) monolithic vs.~(b) hybrid systems~\cite{Wang2021a}, as well as two time snap-shots of the corresponding wavefront depicted by absolute velocity $|\dot{u}|$ [m/s].}
		\label{fig:2}
	\end{figure*}

	In this work we focus on the early-time nonlinear acoustics of the 2D hybrid system subject to 
	a uniform shock distribution applied to the granular medium (see Fig.~\ref{fig:1}b), corresponding to 
	initial velocities of $v_0(0^{+}) = 0.34\,\mathrm{m/s}$ for the left boundary granules and with 
	all other initial conditions set to zero. For comparison, we also consider the linear acoustics 
	of the ``monolithic'' system of Fig~\ref{fig:1}a, composed of the thin ``right plate'' of the hybrid system of Fig.~\ref{fig:1}b, 
	but with the granular assembly replaced by a ``left plate'' of the same composition as the right 
	plate, while preserving the boundary conditions. Moreover, for fair comparison an isoenergetic 
	shock energy intake (i.e., the same energy shock input) is applied to the left boundary of the 
	monolithic system, in the form of a uniform half-sine shock of duration $1\,\mu\mathrm{s}$ and 
	intensity $20{,}000\,\mathrm{N/m}$. Energy measures related to the shock responses of the two 
	material systems are presented in Fig.~\ref{fig:2}, and the accuracy of the simulations is demonstrated 
	by conservation of the total input shock energy when viscoelastic and frictional dissipation 
	effects are accounted for.

The plate is modeled as a linear isotropic medium under plane-stress assumptions.
 The governing equation of elastodynamical motion in the plate is,
	\begin{equation}
		\begin{aligned}
			\rho \ddot{u}_i &= \sigma_{ij,j} +f_{i} \\
			\sigma_{ij} &= c_{ijkl}\varepsilon_{kl},\\
			\varepsilon_{ij} &= \frac{1}{2}(u_{i,j} + u_{j,i}) 
		\end{aligned}
		\label{EQ:StressStrain}
	\end{equation}
	where $\rho$ is density, $\sigma_{ij}$ the stress tensor, $f_i$ the external force, $u_i$ the displacement, $\varepsilon_{ij}$ the strain tensor, and $c_{ijkl}$ the elasticity tensor.
	For monolithic simulations, we consider the domains $\Omega_{\rm L}$ and $\Omega_{\rm R}$ as the left- and right-halves of the plate-plate interface, respectively. 
	For bead-plate simulations, only $\Omega_{\rm R}$ is modeled by Eq~\eqref{EQ:StressStrain}, while a granular medium is modeled in place of $\Omega_{\rm L}$.
	The granular medium is modeled as a 2D hexogonal lattice of spherical granules that obey a Hertzian restoration force. The motion of each granule is dictated by the balance of transverse and angular forces, 
	\begin{equation}
		\begin{aligned}
		\ddot{\boldsymbol{s}}_i &= \frac{1}{m_i}\left(\sum_j (\boldsymbol{N}_{ij}+\boldsymbol{f}_{ij}) + \sum_{k}(\boldsymbol{N}_{ik}+\boldsymbol{f}_{ik}) \right)\\
		\ddot{\bm{\theta}}_i &= \frac{{R}_i}{{I}_i}
		\left(\sum_j (\boldsymbol{n}_{ij}\times \boldsymbol{f}_{ij}) + \sum_{k}(\boldsymbol{n}_{ik}\times \boldsymbol{f}_{ik}) \right)\\
		\end{aligned}
		\label{EQ:Beads},
	\end{equation}
	where $\bm{s}$ is the translational displacement of the granules, $\bm{\theta}$ the angular displacement, $m_i$ is mass, $R_i$ radius, and $I_i$ the moment of inertia, respectively. $\boldsymbol{N}_{ij}$ and $\boldsymbol{n}_{ij}$ denote the normal force and position vectors between granules $i$ and $j$, respectively, while $\boldsymbol{N}_{ik}$ and $\boldsymbol{n}_{ik}$ denote the same quantities with regard to the $i$-th boundary facing granule and the $k$-th point on the boundary of $\Omega_{\rm R}$.
	We consider steel granules with radius 
	$R=0.01\,\mathrm{m}$, Poisson’s ratio $\nu=0.3$, elastic modulus $E=200\,\mathrm{GPa}$, and 
	density $\rho=7850\,\mathrm{N/m^3}$.
	We discretized the elastic continuum using 8-node quadratic elements with a maximum mesh spacing of 1.25 mm, enabling the resolution of wavenumbers up to 2500 rad/m.
	Full details regarding the implementation of Eqs~\eqref{EQ:StressStrain} and~\eqref{EQ:Beads} may be found in Ref~\cite{Wang2021a}.
	
	Considering the early-time wave transmission through the hybrid interface (top of Fig.~\ref{fig:2}), at $t \sim 0.07\,\mathrm{ms}$ 
	the primary wave front (i.e., the wave packet that is initiated immediately after the application 
	of the shock and is not yet ``polluted'' by reflections at the lateral boundaries) reaches the 
	discrete--continuum interface through the 5 contact points (see Fig.~\ref{fig:1}b), and energy starts 
	transmitting to the receiving elastic plate, reaching a maximum of $\sim 41\%$ of the input shock energy 
	at $t \sim 0.12\,\mathrm{ms}$. Following that, a large portion of that energy ``backscatters'' 
	to the granular assembly, as the reflected wave from the right boundary of the plate reaches 
	again the interface, so by the loss of contact between the plate and the boundary granules at 
	the contact points (due to granule--plate separation occurring at $t \sim 0.17\,\mathrm{ms}$) 
	only $\sim 10\%$ of the shock energy is still confined in the receiving plate. A comparison with the 
	linear acoustics of the monolithic elastic plate (Fig.~\ref{fig:2}(a)) reveals clearly the strong nonlinear dispersion of the primary wave front in the hybrid system. Indeed, in the monolithic plate most of the shock energy is captured in the primary wave front throughout the early-time primary wave transmission and gets transmitted and reflected intact back and forward in the form of a coherent, highly energetic, localized wave front (i.e., as a ``block of energy''). 
	This results in a sequence of beat phenomena as shock energy is continuously exchanged between 
	the left (transmitting) and the right (receiving) plates of the monolithic system.

	No such coherent wave front propagation exists in the hybrid system, where the 2D granular 
	assembly inflicts a drastic delay in the primary wave front transmission by ``softening'' 
	the nonlinear acoustics, i.e., by introducing relatively slow time scales in the system 
	response through intense, multi-scale, nonlinear wave dispersion. This time delay is 
	accompanied by drastically diminished energy transfer across the granular--plate interface, 
	i.e., $\sim 41\%$ maximum transmitted shock energy in the hybrid system, compared to an 
	$\sim 88\%$ maximum for the monolithic plate. Moreover, as shown in~\cite{Wang2021a}, 
	the strong nonlinearity of the acoustics of the hybrid system introduces passive tunability 
	of its responses to energy, but this specific feature of the nonlinear acoustics, although 
	interesting, will not be considered further in this work.

	These results highlight the rich potential of hybrid systems for unprecedented multi-scale 
	wave tailoring and shock mitigation. To explore, physically understand, and predictively 
	engineer this class of hybrid systems for wave scattering, one needs to perform a fundamental, 
	in-depth study of the acoustics based on a multi-scale postprocessing framework capable of 
	quantifying the strongly nonlinear wave scattering (dispersion) in the wavenumber and/or 
	frequency domains occurring at the discrete--continuum interfaces. The basic elements of 
	this framework are discussed in the next section.

	\section{Postprocessing framework}\label{SECProcessing}
	
	The principal aim of the postprocessing analysis is the quantification of the multi-scale 
	nonlinear scattering of the primary wavefield that transmits in the elastic plate of the hybrid 
	metamaterial, and its comparison to the respective wavefield that transmits in the right plate 
	of the monolithic system (see Fig.~\ref{fig:2}). As a result, we aim toward gaining an understanding of 
	the strong nonlinear scattering of the transmitted wave by the highly discontinuous granular 
	layer, by comparing it with the ``baseline'' analysis of an isoenergetic wave field in the 
	monolithic plate where the dispersion is less intense (since it transmits through the 
	homogeneous, linearly elastic thin plate).
	
	The postprocessing framework to be followed is based on acoustic wavefield imaging of the 
	primary wavefields and adopts the computational method outlined in~\cite{Flynn2013,Kudela2015,Jeon2017,Jeon2020}. Such 
	wavenumber-domain imaging techniques operate by successively filtering the wavenumber content 
	of propagating or stationary waves to characterize the spectral composition of a wavefield. As such, 
	they are commonly used in the field of nondestructive evaluation (NDE) to quantify local changes 
	in material thickness~\cite{Jeon2020}. At the same time, this methodology may allow for new 
	insights into the effects of wave scattering for nonlinear engineered systems, whereby the 
	spectral spreading of energy is an ultimate goal – e.g., for wave-scattering metamaterials. 
	To the authors’ knowledge, it has not yet been applied to quantify multi-scale scattering of 
	incident waves in material systems with strongly nonlinear internal components such as the 
	hybrid system considered herein.
	
\begin{figure}[t!]
		\centering
		 \includegraphics[width=.5\columnwidth]{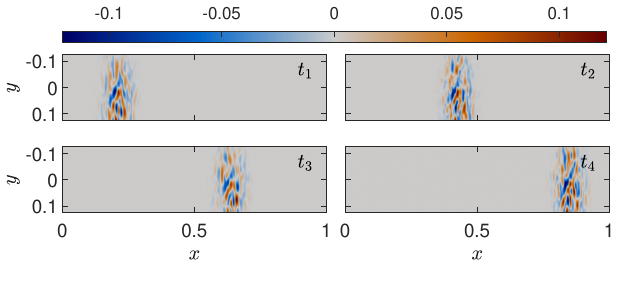} 
		\caption{\justifying Time snapshots of the wavefront for the synthetic wavefield of Eq.~\eqref{eq:wavefield}, with color scales corresponding to scalar-valued displacement of arbitrary units.}
		\label{fig:3}
	\end{figure}

	To demonstrate the postprocessing framework for a propagating wave in a plate, we consider the 
	following 2D rightward propagating synthetic wavefield:
	\begin{equation}
		r(x,y,t) \;=\; \sum_q^Q
		G\bigl(x - c\,t\bigr)\,\exp\!\Bigl\{ i \bigl(k_x\,\delta_{x_q}\,x + k_y\,\delta_{y_q}\,y + \omega\,t \bigr) \Bigr\},
		\label{eq:wavefield}
	\end{equation}
	composed of $Q$ plane waves with random wavenumbers $\delta_{x_q}$ and 
	$\delta_{y_q}$, drawn from a Gaussian distribution, and a Gaussian (spatially localized) envelope $G(\cdot)\in[0,1]$, and velocity 
	$c$. This wavefield is representative of the axial and lateral deformation fields $u_x(x,y,t)$ or 
	$u_y(x,y,t)$, respectively, computed for the hybrid and monolithic systems of Fig.~\ref{fig:2}. Time snapshots 
	of the resulting wavefront are given in Fig.~\ref{fig:3}.
	
	\begin{figure}[t!]
		\centering 
		\includegraphics[width=.5\columnwidth]{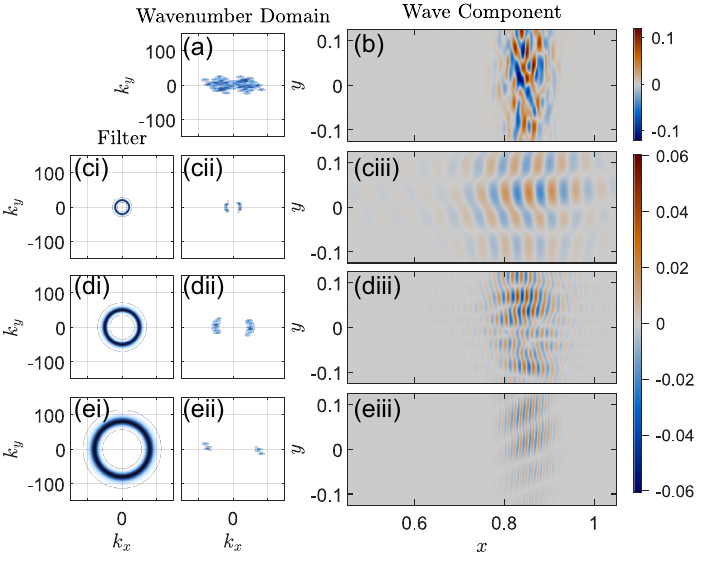}
		\caption{\justifying Postprocessing framework in the 2D wavenumber domain for a snapshot of the synthetic wavefield of Fig.~\ref{fig:3}: 
		(a) 2D FT in space of the original wave front at the specific snapshot $R(k_x,k_y,t)$ and (b) the corresponding wavefield; (c)-(e) The wavenumber domain processesing routine for three selections of $k_c$. Rows  depict (i) the wavenumber domain filter $W(k_x,k_y,k_c)$ for three values of $k_c$, (ii) the product $\hat{R}(k_x,k_y,k_c,t)$, and (iii) the filtered wavefield at each $k_c$.
		}
		\label{fig:ex}
	\end{figure}

	Given a wave snapshot, e.g., at $t = t_4$, we perform a 2D Fourier transform (2D FT)
	in space of the deformation field to obtain the (complex-valued) transformed deformation field 
	\begin{equation}
	R(k_x,k_y,t) \;\equiv\; \mathcal{F}^{(2)}\bigl\{r(x,y)\bigr\}
	\end{equation}
	in the wavenumber domain (Fig.~\ref{fig:ex}(a)), where $k_x$ and $k_y$ are the axial and transverse wavenumbers, 
	respectively which are reported in units of $2\pi/\lambda$ and $\lambda$ is the wavelength in meters. 
	A databank of wavenumber filters, $W(k_x,k_y,k_c)$, is then considered, where $k_c$ 
	is a center wavenumber that controls the radius of the filter band; a larger ring means that higher 
	wavenumber content is kept. Next, the wavenumber map $R(k_x,k_y,t)$ is decomposed by applying a 
	filter bank of expanding radii, yielding a sequence of filtered wavefields at varying $k_c$ as
	\begin{equation}
	\tilde{R}(k_x,k_y,k_c,t) \;=\; R(k_x,k_y,t)\;W(k_x,k_y,k_c).
	\end{equation}
	In the current work, the Log-Gabor wavenumber filter,
	\begin{equation}
	W(k_x,k_y,k_c) \;=\; \exp\!\left[\,
	-\,\frac{\log\!\Bigl(\tfrac{\sqrt{k_x^2 + k_y^2}}{k_c}\Bigr)^{2}}%
	{2\,\log\!\Bigl(\tfrac{\alpha}{k_c}\Bigr)^{2}}\,\right],
\end{equation}
	is used, where $\alpha$ determines the filter spread about $k_c$. This filter is in the form of 
	a ring in the wavenumber domain with radius $k_c$ (Figs.~\ref{fig:ex}(ci-ei)). Each filtered wavenumber field 
	$\tilde{R}(k_x,k_y,k_c,t)$ (Figs.~\ref{fig:ex}(cii-eii)) is then inverse-2D Fourier transformed in space, recovering a filtered 
	wavefield
	\begin{equation}
	\tilde{r}(x,y,k_c,t) \;=\; \bigl[\mathcal{F}^{(2)}\bigr]^{-1}
	\Bigl\{\tilde{R}(k_x,k_y,k_c,t)\Bigr\},
	\end{equation}
	at the selected time instant, by retaining only the wavenumber content close to the selected 
	center wavenumber $k_c$ (Figs.~\ref{fig:ex}(ciii-eiii)). This process isolates and quantifies the component of the 
	propagating wavefield close to each center wavenumber (or wavenumber band) $k_c$ for the specific 
	time snapshot considered.

	Based on this framework, a quantitative multi-scale study of the propagating primary wave front 
	can be performed. 
	Moreover, an energy wavenumber analysis 
	can also be accomplished, since, based on the filtered wavefields $\tilde{r}(x,y,k_c,t)$, the 
	energy of the elastic continuum at different spatial scales (center wavenumbers) $k_c$ can be 
	computed. In that way the nonlinear scattering of the energy of the considered wave front across 
	different spatial scales can be quantified, and the effects of the nonlinear scattering of the 
	transmitted wave by the granular component of the hybrid system can be accurately computed and, hence, fully quantified. 
	Note that similar analysis for the scattering of the kinetic energy of the propagating wave front 
	can be performed based on the computed velocity fields (instead of using deformation fields). 
	Hence, a complete quantification of the dominant energy transfers in the wavenumber domain due 
	to the action of the system nonlinearity will result.
	
	In the next section we apply the outlined postprocessing framework to study the scattering of 
	the primary wave fronts depicted in Fig.~\ref{fig:2}, detect the different patterns of energy transfer 
	in the wavenumber domain for the hybrid and monolithic systems, and, by comparing them, assess 
	the nonlinear wave dispersion in the hybrid system caused by the acoustical interaction of its 
	discrete and continuum components.

	\begin{figure}[t!]
		\centering 
		\begin{subfigure}{.5\linewidth}
			\includegraphics[width=\columnwidth]{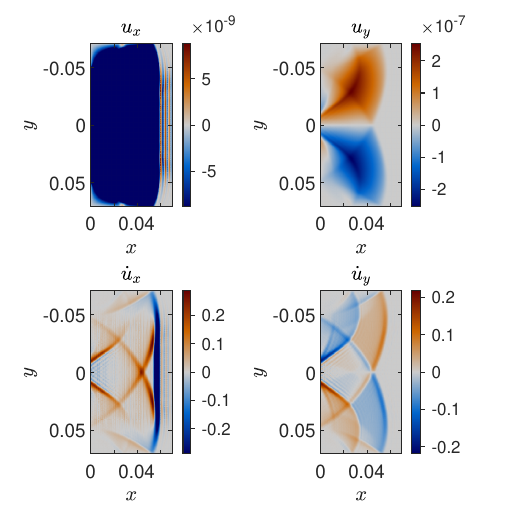} 
			\caption{Monolithic Interface}
		\end{subfigure}%
		\begin{subfigure}{.5\linewidth}
			\includegraphics[width=\columnwidth]{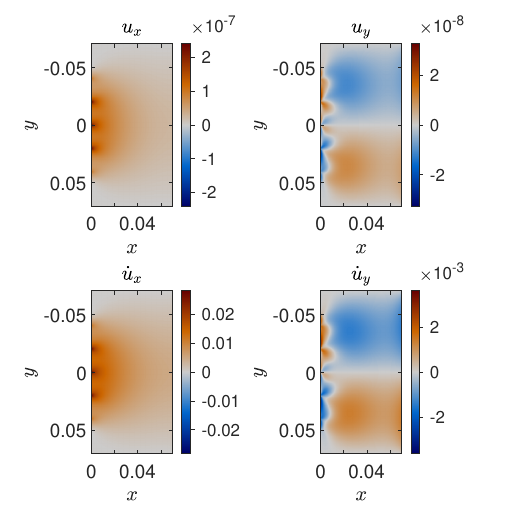} 
			\caption{Granular Interface}
		\end{subfigure}
		\caption{\justifying Displacement and velocity fields of propagating wave front for the (a) monolithic and (b) granular interfaces corresponding to the time snapshots of Fig.~\ref{fig:2}(a) for $t=2.05$e$-05$, and Fig.~\ref{fig:2}(b) for $t$=8.7e$-05$, respectively. 
			Color scale units are given in m and m/s for displacement ($u_x$ and $u_y$) and velocity ($\dot{u}_x$ and $\dot{u}_y$), receptively. }
		\label{fig:velocity}
	\end{figure}
	
	\section{Energy scattering in the wavenumber domain}\label{SECresults}
	
\begin{figure}[t!]
		\centering 
		\begin{subfigure}{.5\linewidth}
			\includegraphics[width=\columnwidth]{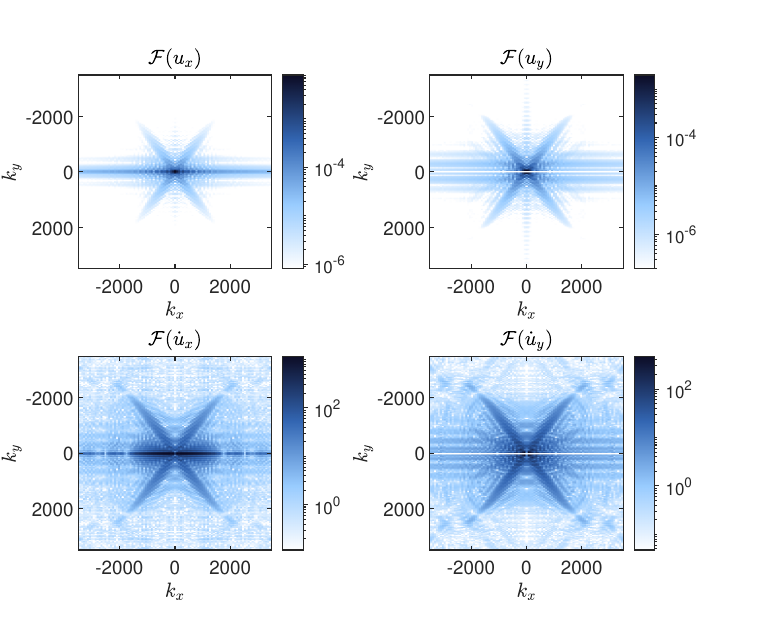}  
			\caption{Monolithic Interface}
		\end{subfigure}%
		\begin{subfigure}{.5\linewidth}
			\includegraphics[width=\columnwidth]{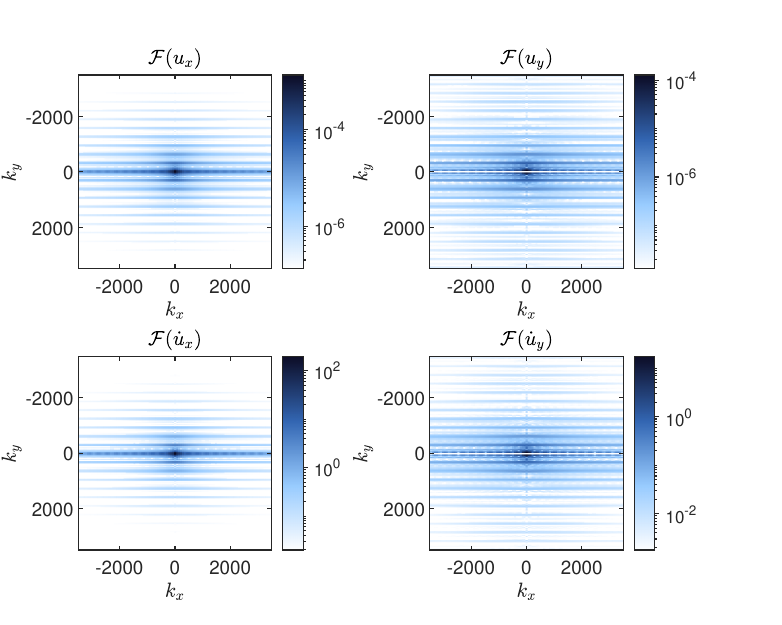} 
			\caption{Granular Interface}
		\end{subfigure}
		\caption{\justifying Spectral analysis (2D FT) of the snapshots of Fig.~\ref{fig:velocity}.} 
		\label{fig:Fdomain}
	\end{figure}

	As a first demonstration of the postprocessing methodology we provide the 2D wavenumber 
	transformation to the two representative time snapshots of the transmitted wave in the 
	right plate of the baseline monolithic system at $t=2.05$e$-05$ seconds (Figs.~\ref{fig:velocity}(a)), and the scattered wave in the elastic 
	plate of the hybrid system at $t=8.7$e$-05$ s (Fig.~\ref{fig:velocity}(b)). To apply the wavenumber processing scheme described in 
	Section~3, the results of the FE simulations were first interpolated over a uniform grid in 
	$x$ and $y$ and padded to reduce edge effects. In Fig.~\ref{fig:Fdomain} we depict the corresponding spectral 
	analysis in the 2D wavenumber domain, that is, the contour plots of the 2D FT of these snapshots. 
	The difference in the spectral decompositions is apparent and reflects the difference between 
	the wave-like transmission in the baseline monolithic system, compared to the diffusion-like 
	transmission in the plate of the hybrid system. Specifically, in contrast to the radial spreading 
	of energy in the $(k_x,k_y)$ wavenumber domain for the baseline system, there is localization of 
	energy mainly along the line $(k_x,k_y = 0)$ for the hybrid system; this is particularly profound 
	for the $x$-components of the wavefield in Fig.~\ref{fig:velocity}(b), indicating minimal propagation in the 
	$y$-direction of the scattered wavefield. Overall, we observe the drastic change in the 
	spectral composition of the propagating wavefront inflicted by its strongly nonlinear wave scattering 
	in the granular part of the hybrid system.

	
\begin{figure}[t!]
		\begin{subfigure}{.5\linewidth}
			\includegraphics[width=\linewidth]{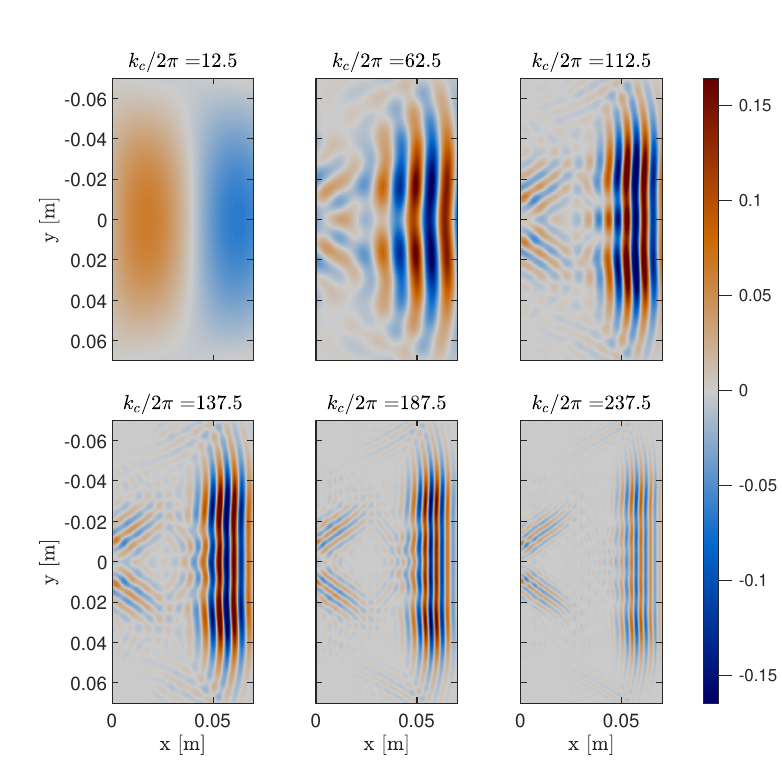}
			\caption{$\dot{u}_x$}
		\end{subfigure}%
		\begin{subfigure}{.5\linewidth}
			\includegraphics[width=\linewidth]{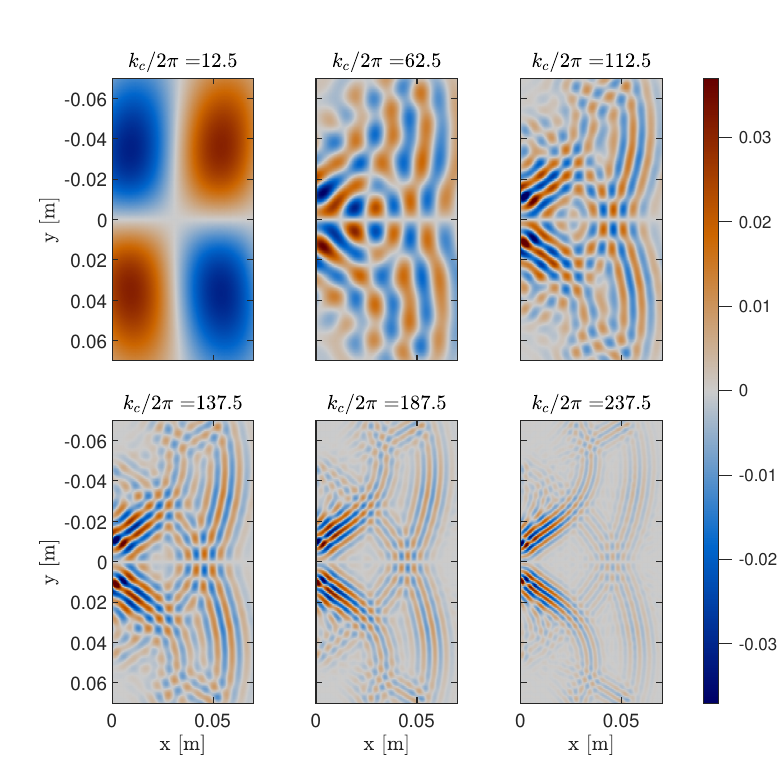}
			\caption{$\dot{u}_y$}
		\end{subfigure}
		\caption{\justifying   Wavefield reconstruction of the $x$- and $y$-velocity snapshots in the receiving plate of 
			the baseline monolithic system of Fig.~\ref{fig:velocity}(a) at varying center wavenumber $k_c$ (shown at top of 
			each reconstruction).}
		\label{fig:rec_pp}
	\end{figure}

\begin{figure}[t!]
	\begin{subfigure}{.5\linewidth}
		\includegraphics[width=\linewidth]{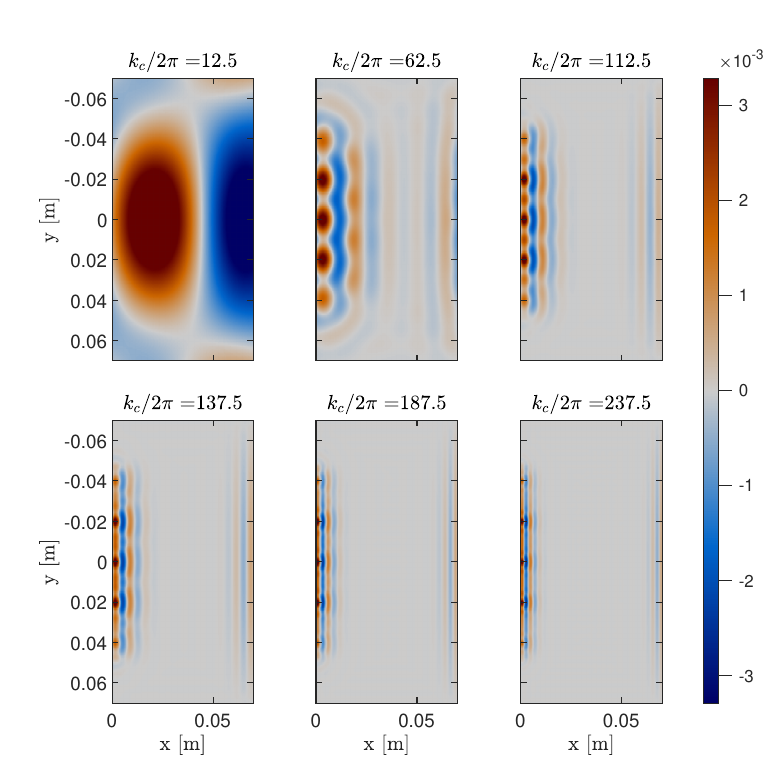}
		\caption{$\dot{u}_x$}
	\end{subfigure}%
	\begin{subfigure}{.5\linewidth}
		\includegraphics[width=\linewidth]{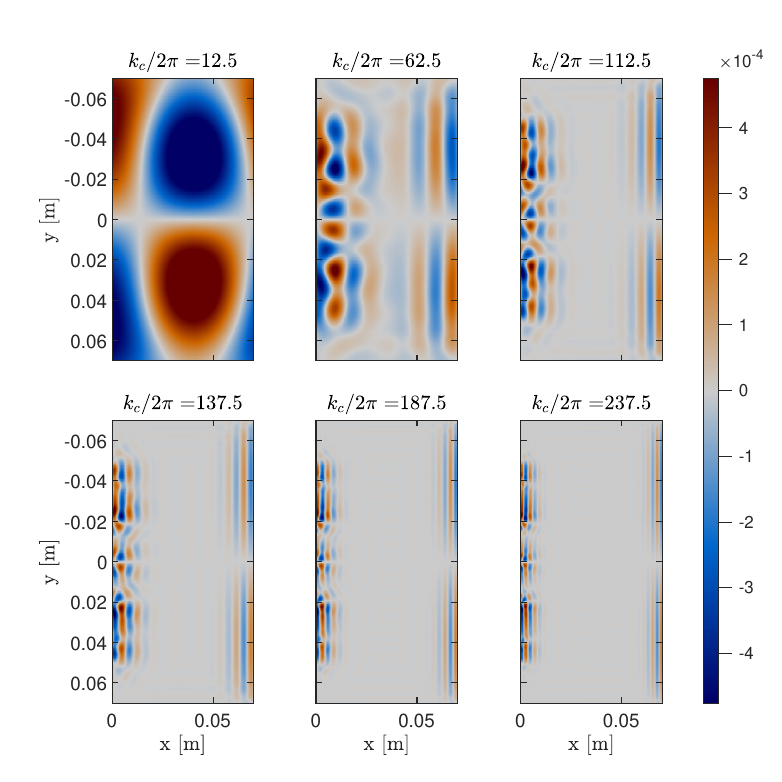}
		\caption{$\dot{u}_y$}
	\end{subfigure}
	\caption{\justifying Wavefield reconstruction of the $x$- and $y$-velocity snapshots in the receiving plate of  the 
		hybrid system of Fig.~\ref{fig:velocity}(b) at varying center wavenumber $k_c$ (shown at top of each reconstruction).}
	\label{fig:rec_bp}
\end{figure}

	Proceeding to full post-processing of the transmitted wavefie``lds, in Figs.~\ref{fig:rec_pp} and~\ref{fig:rec_bp} we reconstruct 
	the $x$- and $y$-velocity snapshots of the baseline and hybrid systems considered in Fig.~\ref{fig:velocity} at varying 
	center wavenumbers $k_c/2\pi \in [25,50,75,100,125,150,175,200,225]$. In essence, these results 
	represent spectral representations of wave scattering in the 2D wavenumber domain.

\begin{figure}[t!] \centering
		\includegraphics[width=.5\linewidth]{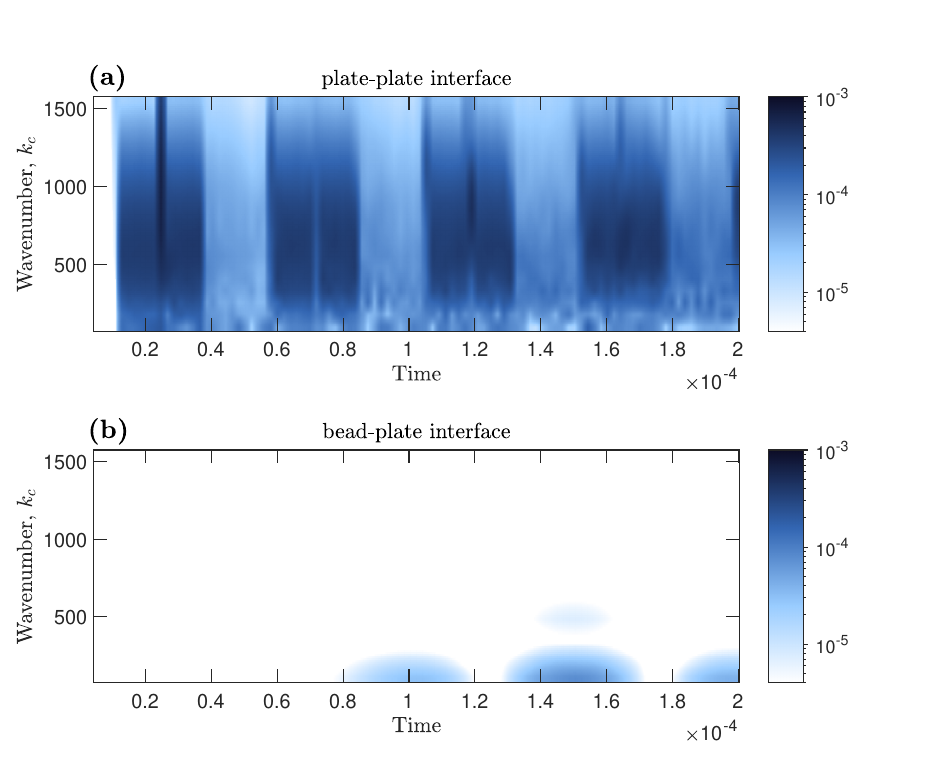}
		\caption{\justifying Kinetic energy scattering versus center wavenumber $k_c$ and time:
			(a) Right plate of the monolithic system, (b) plate in hybrid system.}
		\label{fig:Energy}
	\end{figure}

	These results provide the detailed spectral decomposition of the transmitted wavefields of the receiving plate at 
	various center wavenumbers. As shown below, these decompositions allow us to compute an 
	integrated picture of the scattering of the energy of the transmitted wave in the wavenumber 
	domain and quantify the strong nonlinear dispersion of the 2D propagating wave by the granular 
	layer. The plots of Fig.~\ref{fig:Energy} provide a numerical summary of the spectral energy composition with 
	respect to time. By applying the wavenumber decompositions of Figs.~\ref{fig:rec_pp} and ~\ref{fig:rec_bp}, the kinetic energy 
	distribution of each spectral band may be computed by resampling the numerical data and 
	computing the kinetic energy of the plate for each center wavenumber $k_c$ using the expression
	\begin{equation}
	T(k_c,t) \;=\; \tfrac{1}{2}\,\dot{\textbf{x}}_{k_c}(t)^\intercal\,\textbf{M}\,\dot{\textbf{x}}_{k_c}(t),
	\end{equation}
	where $\dot{\textbf{x}}_{k_c}(t)$ are the sequence of discretized velocity snapshots for varying $k_c$ shown in Figs.~\ref{fig:rec_pp} and ~\ref{fig:rec_bp}, and 
	$\textbf{M}$ is the resampled mass matrix of the plate at the grids used in these figures. In the 
	receiving plate of the hybrid system during primary wave transmission (i.e., for 
	the early-time nonlinear acoustics), these results partially correlate with the plate energies 
	depicted in Fig.~\ref{fig:2}(b) for the same systems, showing the percentages of the total energy input 
	transmitted in the plate including both kinetic and potential energy terms, as well as snap-shots of the absolute velocity field $|\dot{u}|$.

	In addition, these results clearly show the difference in the time scales of the early acoustics 
	for the baseline and hybrid interfaces. Indeed, in the early time-interval considered in Fig.~\ref{fig:Energy}(a), 
	there occur repeated reflections of the transmitted wavepacket at the right boundary of the 
	baseline receiving plate; compared to the plot of Fig.~\ref{fig:Energy}(b), there occur much fewer reflections in the plate 
	of the receiving plate of the hybrid system. More importantly, comparing the two results we notice ``spreading'' of wave 
	energy over the entire wavenumber domain in the baseline plate -- which is representative of 
	transmission, reflection, and interference of spatially extended transient waveforms in 2D elastic bounded media, forming a coherent (compact) wavefront.
	This wavenumber spreading directly corresponds to the coherent (un-scattered) wavefront  
	propagating to the receiver plate in the monolithic study. Moreover, the broad-band spectral composition of the monolithic interface may be expected from the Fourier uncertainty principle, since a (relatively) high-variance spectral composition is required to support localized coherent propagation.

	On the contrary, concerning the receiving plate of the hybrid interface we detect localization of kinetic 
	energy mainly to the smaller wavenumbers, which signifies the presence of larger wavelengths in 
	the transmitted wavefront; this is a feature that is not characteristic of transient wavefront 
	transmission in linearly elastic media, where in early-time acoustics one expects the strong 
	participation of smaller wavelengths, i.e., of high-frequency oscillatory wavepackets, which 
	are required to support localized wavefronts (as in the monolithic case - see comments above).. This effect is a result of the energy spreading 
	and strong wave dispersion induced by the nonlinear granular layer as the primary wave propagates through it, before it encounters the left boundary of the receiving plate (at the interface). Namely, the spectral content 
	of the wave imparted onto the receiving plate for the hybrid system indicates a gradual diffusive energy 
	transfer, whereby low-frequency (near-zero frequency) components dominate the spatial spectrum, 
	and larger bending modes, which are excited gradually through the diffusive energy transfer 
	transmitted by the granular medium, preside over the high-frequency modes required for localized 
	propagation.
	
	Apart from ``slowing'' the early-time nonlinear acoustics in the hybrid interface, as wave components with larger wavenumbers propagate at smaller phase (and group) velocities compared to wave components with smaller wavenumbers, these findings highlight the drastic 
	change in the transmitted wavefront caused by the nonlinear dispersion of the applied shock by 
	the granular layer~\cite{Wang2021a}, which results in non-oscillatory and diffusion-like 
	wave transmission in the receiving plate. Lastly, we notice the drastically reduced levels of 
	kinetic energy of the plate of the hybrid system, indicating the enhanced shock mitigation 
	achieved due to the dispersion of the transmitted shock through the granular layer.

	To  provide further clarity on the above discussion, the results of the spectral decomposition shown in Figs~\ref{fig:rec_pp} and~\ref{fig:rec_bp} were reproduced for a selection of time snap-shots between 4.5 and 200 $\mu$s. 
	The results are located in the supplemental material~\cite{SM} as Figs.~S1 and~S2 for the monolithic and granular interfaces, respectively.
	We refer the reader to the supplemental material and briefly comment here that a relative lack of high-wavenumber content can be observed for the granular interface as compared to the monolithic interface.
	
	
	Commenting on the utility of this analysis, it is worth noting that the method offers a quantitative link between the properties of a propagating medium (e.g., the granular layer versus the monolithic plate) and the spectral composition of transferred wave energy. While the present work serves primarily to establish this framework, future efforts could build on these findings to design scattering media that transmit tailored spectral content through inverse modeling approaches. This capability may be particularly useful in the context of passive nonlinear energy management for structural systems or acoustic waveguides under shock and vibration loading.
	\rev{Moreover, by enabling a mode-resolved view of energy transport, the analysis naturally lends itself to comparison with classical diffusive processes—an analogy we explore in the following section.}

	\section{Analogy to Diffusive Processes}
	\label{SEC:Diff}\rev{
Here we extend the spectral analysis to compare the wave scattering observed in the granular simulation to that of a classical heat diffusion problem. Although the governing physics differ fundamentally, e.g., hyperbolic wave propagation versus parabolic diffusion, the qualitative behavior of energy flow in the elastodynamic domain closely emulates a diffusive, rather than a purely elastic, process. While analogies between elastic scattering and diffusion have been drawn before---particularly in the context of radiative transport modeling for isotropic and linear scatterers~\cite{stam1995multiple,jia2004codalike,mayor2014sensitivity,fuqiang2022progress,van1999multiple}---such comparisons typically rely on bulk transport measures.
In contrast, we focus on the spectral characteristics within the receiving elastic domain itself, instead of within the granular (scattering) layer. 
For clarity and interpretability in the following analysis, we isolate the characteristics of the primary wavefront and neglect boundary reflections.
By tracking the spectral evolution of the transmitted primary wavefront in the elastic medium, we achieve a mode-resolved characterization of the diffusive-like transport enabled by granular scattering. This can be subsequently calibrated against a heat diffusion model and directly contrasted with coherent wave propagation through a monolithic interface (see Fig~\ref{fig:2}(a)).
}

\begin{figure}[t!]
	\centering
	\includegraphics[width=\linewidth]{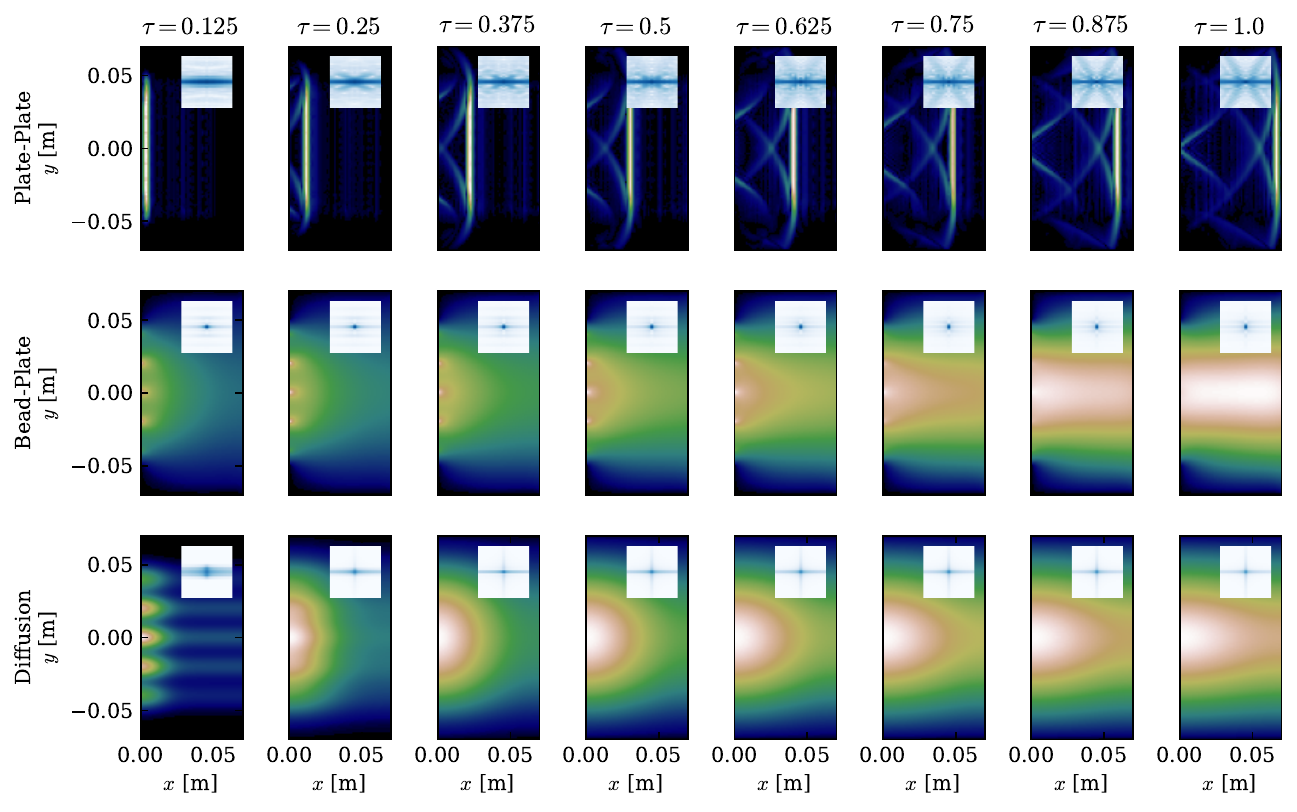}
	\caption{Comparison of propagating energy over normalized time snapshots $\tau\in[0,1]$ for the elastic monolithic, elastic granular, and heat-diffusion problems. The normalized color schemes depict $|\dot{\bm{u}}|$ for the elastodynamic domains and $\psi$ for the head problem. The upper-right inset of each subplot depicts the corresponding wavenumber spectrum for $k_x,k_y\in[-2500,2500]$.}
	\label{fig:propagation}
\end{figure}

\rev{
We consider a baseline heat-diffusion process governed by the classical anisotropic heat equation, subject to free-fixed boundary conditions that emulate the elastic models of Fig.~\ref{fig:1}. To mimic the energy transmitted through the interface beads depicted by Fig~\ref{fig:2}(b), we consider a multi-modal
 Gaussian initial condition designed qualitatively match the early-time responses between the elastodynamic and diffusion problems. The resulting heat-diffusion problem reads,
\begin{align}
	&\psi^\prime = \alpha_x \psi_{xx} + \alpha_y \psi_{yy}, \quad (x, y) \in \Omega_{\text{R}} \label{EQ:Diffusion} \\
	&\psi(0, y, \tau) = \psi(L, y, \tau) = 0,\quad \psi_{y}(x, 0, \tau) = \psi_{y}(x, H, \tau) = 0 \\
	&\psi(x, y, 0) = \sum_k M_k \exp\left( -\frac{1}{2} (\bm{x} - \bm{\mu}_k)^\intercal \Sigma^{-1} (\bm{x} - \bm{\mu}_k) \right) \label{EQ:IC}
\end{align}
Here, $\alpha_x$ and $\alpha_y$ are the diffusion coefficients in the $x$ and $y$ directions, respectively, $\psi$ denotes the heat field, $\tau\in[0,1]$ is normalized time, and $M_k$ ($k = 1, 2, \dots, 5$) are the amplitudes of the Gaussian intruders centered at $\bm{\mu}_k$ and parameterized with diagonal covariance. The notation $(\square)^\prime$ denotes a derivative with respect to $\tau$.
We take $\alpha_x = 2\alpha_y$ to better approximate the anisotropic energy spreading observed in the elastodynamic system. The numerical values of the coefficients are selected arbitrarily since the solutions are considered on a normalized time scale such that energy reaches the right boundary at $\tau=1$, as explained in the subsequent paragraph; here, we only wish to provide qualitative comparisons.
Solutions to Eqs.~\eqref{EQ:Diffusion}–\eqref{EQ:IC} are recovered via the standard separation of variables approach:
\begin{align}
	u(x, y, \tau) &= \sum_{m=1}^{\infty} \sum_{n=1}^{\infty} A_{mn}, 
	\cos(k_x(m) x) \sin(k_y(n) y) \, e^{-\lambda_{mn} \tau}, \label{EQ:HeatSol} \\
	\lambda_{mn} &= \alpha_x k_x^2(m) + \alpha_y k_y^2(n), \label{EQ:Lambda} \\
	A_{mn} &= \frac{4}{LH} \int_0^H \int_0^L \psi(x, y, 0) 
	\cos(k_x(m) x) \sin(k_y(n) y) \, dx\,dy \label{EQ:Amn}
\end{align}
where $k_x(m) = \frac{m\pi}{L}$ and $k_y(n) = \frac{n\pi}{H}$, with $L$ and $H$ denoting plate width and height, respectively. The coefficients $A_{mn}$ provide analytically derived spectral measures of the diffusion process, which can be used to calibrate subsequent FFT-based analyses.
}

\rev{
The propagation of energy in the two elastic systems of Fig.~\ref{fig:1} and the heat diffusion model is illustrated in Fig.~\ref{fig:propagation}, presenting the normalized velocity fields for the elastodynamic models and normalized temperature field for the heat diffusion case, respectively.
The time axis for each simulation is normalized such that $\tau \in [0,1]$, where $\tau = 0$ marks the moment the primary wavefront enters the domain, and $\tau = 1$  the time when it reaches the right boundary. A qualitative inspection of the spatial and spectral wavefields suggests strong visual similarity between the granular elastodynamic simulation and the heat diffusion solution. Whereas the monolithic elastic system maintains a coherent, localized propagating wavefront, the granular interface scatters energy into a more spatially diffuse front. Correspondingly, the wavenumber spectrum of the granular system collapses toward a distribution dominated by low-frequency modes, closely resembling the spectral behavior of the diffusion process.}
 
\rev{
To analyze spectral energy redistribution, we utilize two metrics that track the temporal evolution of the spectral content: spectral variance ${\sigma}_k(\tau)$ and spectral entropy $H_k(\tau)$. Spectral variance is a second-moment measure used to quantify the decay of high-frequency content over time, while entropy captures the overall complexity or disorder of the spectral distribution. 
Combined, these serve as proxies for spectral spreading and enable a direct comparison between the responses of the granular-plate wave system and the diffusion model.
To evaluate these quantities in the elastodynamic simulations, we compute empirical spectral measures $\hat{\sigma}_k(\tau)$ and $\hat{H}_k(\tau)$ using a spatial Fourier representation based on the discrete cosine transform (DCT) in $x$ and discrete sine transform (DST) in $y$. This choice of basis reflects the boundary conditions imposed on $\Omega_{\text{R}}$ and ensures consistency with the modal basis $\{A_{mn}\}$ obtained in the heat solution. Defining $\hat{U}_{mn}(t) = \mathcal{F}_{\text{DST}_y} \circ \mathcal{F}_{\text{DCT}_x} \left[ | \bm{\dot{u}}(\tau) | \right]$, we compute:
\begin{align}
	\hat{\sigma}_k^2(\tau) &= \frac{\sum_{m,n} \left( k_x^2(m) + k_y^2(n) \right) |\hat{U}_{mn}(\tau)|^2}{\sum_{m,n} |\hat{U}_{mn}(\tau)|^2}, \label{EQ:Smn} \\
	\hat{H}_k(\tau) &= - \sum_{m,n} \hat{P}_{mn}(\tau) \log \hat{P}_{mn}(\tau), \label{EQ:Hmn} \\
	\hat{P}_{mn}(\tau) &= \frac{|\hat{U}_{mn}(\tau)|^2}{\sum_{m,n} |\hat{U}_{mn}(\tau)|^2}. \label{EQ:Pmn}
\end{align}
For the closed-form heat diffusion solution, we compute the analytical spectral variance $\sigma_k^2(\tau)$ and entropy $H_k(\tau)$ analytically by substituting $\hat{U}_{mn}(\tau)$ with $A_{mn} e^{-\lambda_{mn} \tau}$ into Eqs.~\eqref{EQ:Smn}–\eqref{EQ:Pmn}. This provides a benchmark for comparison against empirical, simulation-derived quantities and is referred to as the analytical diffusion solution.
}

\begin{figure}\centering
	\begin{subfigure}{.5\linewidth}
		\includegraphics[width=\linewidth]{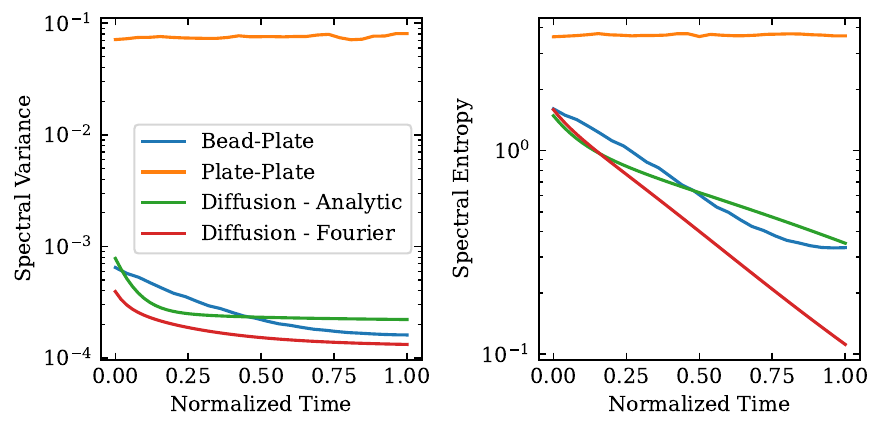}
		\caption{Spectral Variance and Entropy}
	\end{subfigure}%
	\begin{subfigure}{.5\linewidth}
		\includegraphics[width=\linewidth]{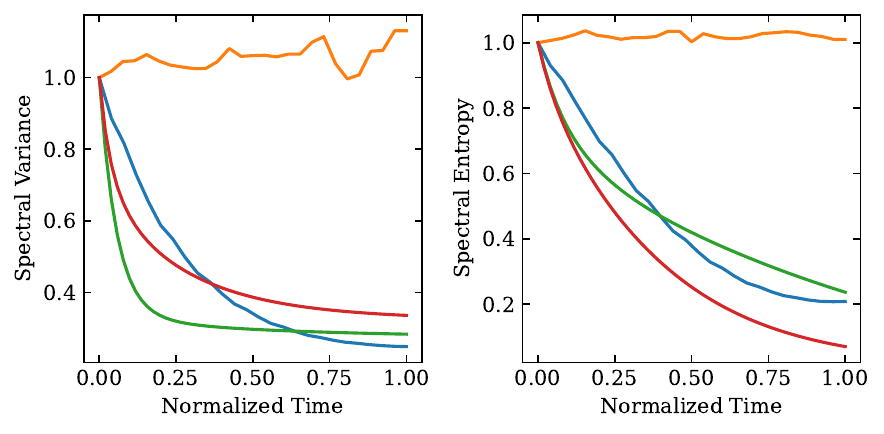}
		\caption{Normalized Spectral Variance and Entropy}
	\end{subfigure}
	\caption{The comparison of (a)  un-normalized and (b) normalized spectral entropy and variance between the elastic systems and diffusion process. The diffusion process depicts both the analytical measures recovered from the separation of variables solution and empirical Fourier based mesasures.}
	\label{Fig:SpectralComp}
\end{figure}	

\rev{
Figure~\ref{Fig:SpectralComp} shows the spectral variance and entropy for all three systems. In Fig.~\ref{Fig:SpectralComp}(a), absolute values are plotted per Eqs.~\eqref{EQ:Smn}–\eqref{EQ:Pmn}, while Fig.~\ref{Fig:SpectralComp}(b) displays values normalized by their initial state at $\tau = 0$. This normalization allows for clearer visual comparison of spectral evolution over $\tau \in [0, 1]$.
In the monolithic plate case, both spectral entropy and variance show minimal change, with values remaining significantly higher than those observed in the granular or diffusion systems. This confirms that the wavefront remains coherent and highly localized throughout its propagation, consistent with the visual snapshots of the wavefields.
In contrast, the granular and diffusion cases both exhibit low initial entropy and variance (note the log scale in Fig.~\ref{Fig:SpectralComp}(a)) and show clear temporal evolution. As $\tau$ increases, both metrics decrease steadily, reflecting a loss of coherence and redistribution of energy across spatial modes. Notably, both systems follow a similar decay trend, roughly proportional to $1/\tau^2$, which aligns with the expected behavior of the diffusion solution from Eq.~\eqref{EQ:HeatSol}.}

\rev{
The trends of Fig~\ref{Fig:SpectralComp}(a) are further corroborated by
Fig~\ref{Fig:SpectralComp}(b) which shows that the normalized monolithic spectral characteristics follow no meaningful temporal evolution, whereas the granular system aligns closely to the monolithic decay reported for both the analytical and Fourier-based characteristics computed for the diffusion process.
We note that at times the spectral characteristics of the granular and diffusion processes diverge from one-another.
This is due to the conservative nature of elastic wave scattering, where total energy remains constant, but directional coherence is lost. 
Nevertheless, the normalized measures clarify  the prominent differences between the monolithic and granular propagation, emphasizing the temporal similarity in decay trends between the granular and diffusion models while clearly separating them from the invariant response of the plate–plate system. 
\textcolor{black}{Moreover, the Fourier-based diffusion metrics align closely with the analytical separation-of-variables solution, reinforcing the reliability of the spectral diagnostics. Minor differences between the two approaches can be attributed to discrepancies in spectral resolution: the FFT-based method operates on a finer spatial grid and captures higher-frequency content beyond the truncated modal expansion used for computing $A_{mn}(t)$, which can influence both entropy and variance estimates.}
Hence, the spectral analysis formulated herein demonstrates that the granular interface reveals clearly the granular-caused diffusion-like behavior in the elastodynamic domain, in clear contrast to the coherent response of the monolithic system.
}

	\section{Concluding remarks}\label{SECConclusion}
	
	We developed a computational method for quantifying the nonlinear multi-scale scattering of a 
	wavefront transmitted across a 2D granular--elastic plate interface. This was achieved through 
	acoustic wavefield imaging, which yielded a detailed quantification of the scattering of the 
	energy of the wave in the wavenumber domain. In a more general context, the method enables 
	implementation of intentional nonlinearity in practical hybrid material systems, i.e., in 
	systems composed of continuum and discrete interacting parts, since it provides predictive 
	capacity for assessing the capacity of the nonlinearity to scatter (direct) wave energy over different  (and even a priori designated favorable)
	wavenumber bands. Based on this information, for example, one could perform inverse design to 
	predictively implement a certain form of (even strong) nonlinearity to tailor the wave energy 
	scattering across an interface in preferred wavenumber bands, while avoiding other wavenumber 
	bands.
	\rev{In addition, we demonstrated that the spectral evolution in the transmitted primary wavefront induced by the nonlinear granular interface resembles a diffusive process in the wavenumber domain. This emergent behavior—typically associated with linear multiple scattering theories—was shown to arise here from intense, strongly nonlinear wave dispersion in a compact region. As such, the granular interface functions as a physical analog of a diffusive scattering layer, offering new insights into the transport of wave energy in hybrid nonlinear systems.}
	
	To the authors’ knowledge, the outlined method applies for the first time to this type of acoustic 
	wavenumber processing, which was originally developed in the field of non-destructive 
	evaluation, to study the energy scattering of waves propagating in acoustic metamaterials with 
	discrete--continuum interfaces. Hence, the method can find application in exploring the 
   acoustics of linear elastic media with embedded nonlinear layers or local nonlinear inclusions.


\bibliographystyle{elsarticle-num}
	\bibliography{Refs}
	\newpage 	

	\clearpage
\begin{center}
	\textbf{\large Supplemental Material}\\
\end{center}

\renewcommand{\theequation}{S\arabic{equation}}
\renewcommand{\thefigure}{S\arabic{figure}}
\renewcommand{\thetable}{S\arabic{table}}
\setcounter{equation}{0}
\setcounter{figure}{0}
\setcounter{table}{0}

\begin{figure*}[h!]\centering
	\includegraphics[width=\linewidth]{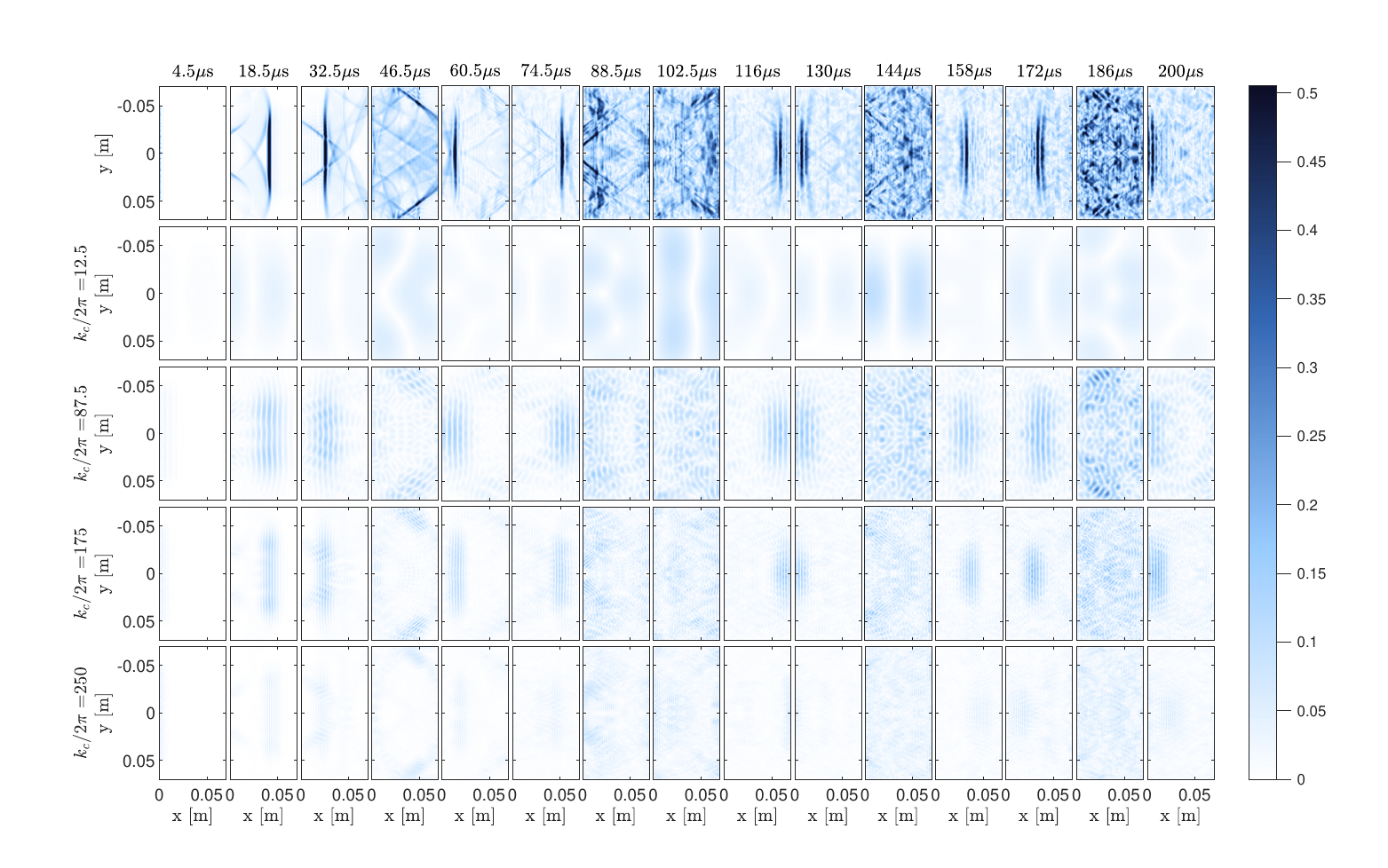}
	\caption{\justifying The demonstrated decomposition  for various time snap-shots of the wave propagation in the monolithic system. The top row depicts the total simulation velocity, $|\dot{u}|$, while the next four rows depict its spectral decomposition at specified central wavenumber bands.}
	\label{Fig:S1}
\end{figure*}	
\begin{figure*}[h!]\centering
	\includegraphics[width=\linewidth]{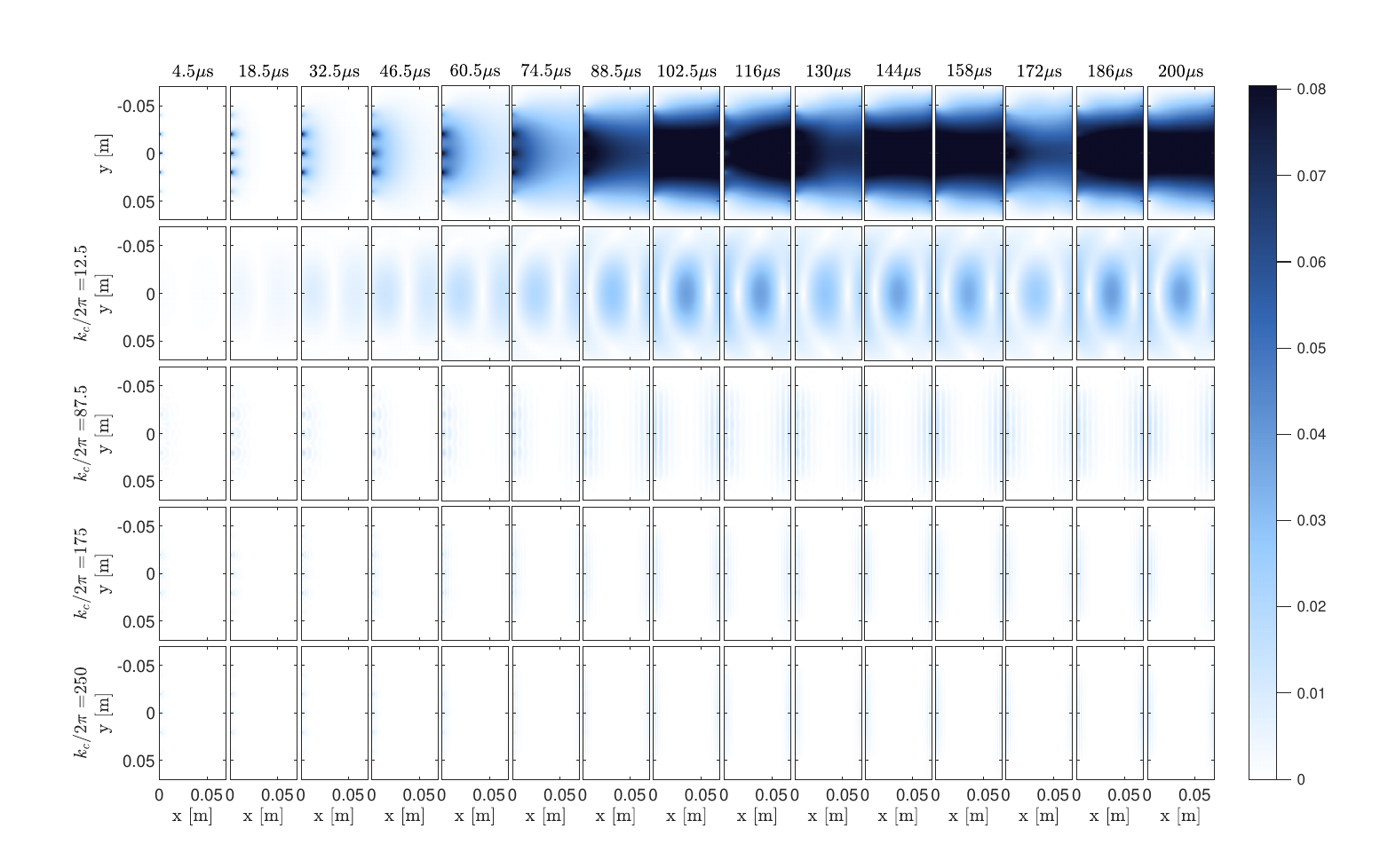}	\caption{\justifying The demonstrated decomposition  for various time snap-shots of the wave propagation in the granular system. The top row depicts the total simulation velocity, $|\dot{u}|$, while the next four rows depict its spectral decomposition at specified central wavenumber bands.}
	\label{Fig:S2}
\end{figure*}

\end{document}